\documentclass[twocolumn,preprintnumbers,amsmath,amssymb]{revtex4}

\usepackage{graphicx}
\usepackage{dcolumn}
\usepackage{bm}

\begin{document}


\title{Supersymmetry Breaking Casimir Warp Drive}

\author{Richard K Obousy}

\email{Richard\_K\_Obousy@baylor.edu}
\affiliation{%
Baylor University, Waco, Texas, 76706, USA
}%

\date{\today}

\begin{abstract}
Within the framework of brane-world models it is possible to account
for the cosmological constant by assuming supersymmetry is broken on
the 3-brane but preserved in the bulk. An effective Casimir energy
is induced on the brane due to the boundary conditions imposed on
the compactified extra dimensions. It will be demonstrated that
modification of these boundary conditions allows a spacecraft to
travel at any desired speed due to a local adjustment of the
cosmological constant which effectively contracts/expands space-time
in the front/rear of the ship resulting in motion potentially faster
than the speed of light as seen by observers outside the
disturbance.
\end{abstract}

\maketitle

\section{Introduction}

Over the last decade there has been theoretical interest in
curiosities dubbed 'warp-drives' initiated by the 1994 paper by M.
Alcubierre [1]. These warp drives are constructs that allow some
object (a spacecraft) to travel at superluminal velocities by
manipulating spacetime in a way such that the spacecraft never
locally exceeds the speed of light, but in a manner identical to the
inflationary stage of the universe the spacecraft does have a
relative speed defined as change of proper spatial distance over
proper spatial time faster than the speed of light.

Interest in warp drives has not been solely confined to the realm of
theoretical speculation as shown by the formation of the NASA
Breakthrough Propulsion Program [2] and the British Aerospace
Project Greenglow [3] both of whose purpose has been to investigate
the realization of these ideas.

In the spirit of the Morris, Thorne and Yurtsever paper [4] these
warp drives, thought highly speculative in nature, provide an unique
and inspiring opportunity to ask the question 'what constraints do
the laws of physics place on the abilities of an arbitrarily
advanced civilization'. In this paper a new and  innovative
mechanism to generate the necessary 'Alcubierre warp bubble' is
proposed.

It has been suggested in the context of brane-world models that our
universe is a (3+1) brane residing in some higher dimensional bulk
[5]. It is known Phenomenologically that supersymmetry is broken on
our 3-brane, however is has been suggested that it may not be broken
on the bulk [6].

\begin{figure}
\includegraphics[width=3.25in]{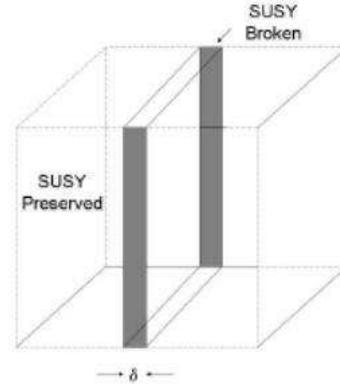}
\caption{Representation of three brane with thickness $\delta$
existing in higher dimensional bulk.}
\end{figure}

Unbroken SUSY decrees that the components of the Chiral or Gauge
Multiplets share equal masses in the bulk and have the same
interaction strength, however on the 3-brane SUSY breaking induces a
mass square difference between them. Motivated by string theory the
3-brane has an effective thickness $\delta$ characterized by the
string thickness $l_s$. As a result the Casimir energy is
non-trivial in the extra dimensional volume that encompasses the
brane. This energy has the necessary features to account for the
cosmological constant.

For simplicity assume an $M^4 \otimes T^n$ manifold with extra
dimensional radius a. SUSY breaking around the brane alters the
Casimir energy which leads to a mass shift of the bulk fields. It is
the aim of this paper to demonstrate that the mass shift is directly
related to the radius of the extra dimension and as such a local
change in the radius of the extra dimension will have the effect of
altering the mass shift and thus the Casimir Energy which locally
effects the value of the cosmological constant in the region
effectively creating a 'bubble' of inflation/contraction.

A spacecraft with the ability to create such a bubble will always
move inside their own local light-cone however the ship can utilize
the expansion of spacetime behind the ship to move away from some
object at any desired speed or equivalently to  contract the
spacetime in front of the ship to approach any object.

\begin{figure}
\includegraphics[width=2.50in]{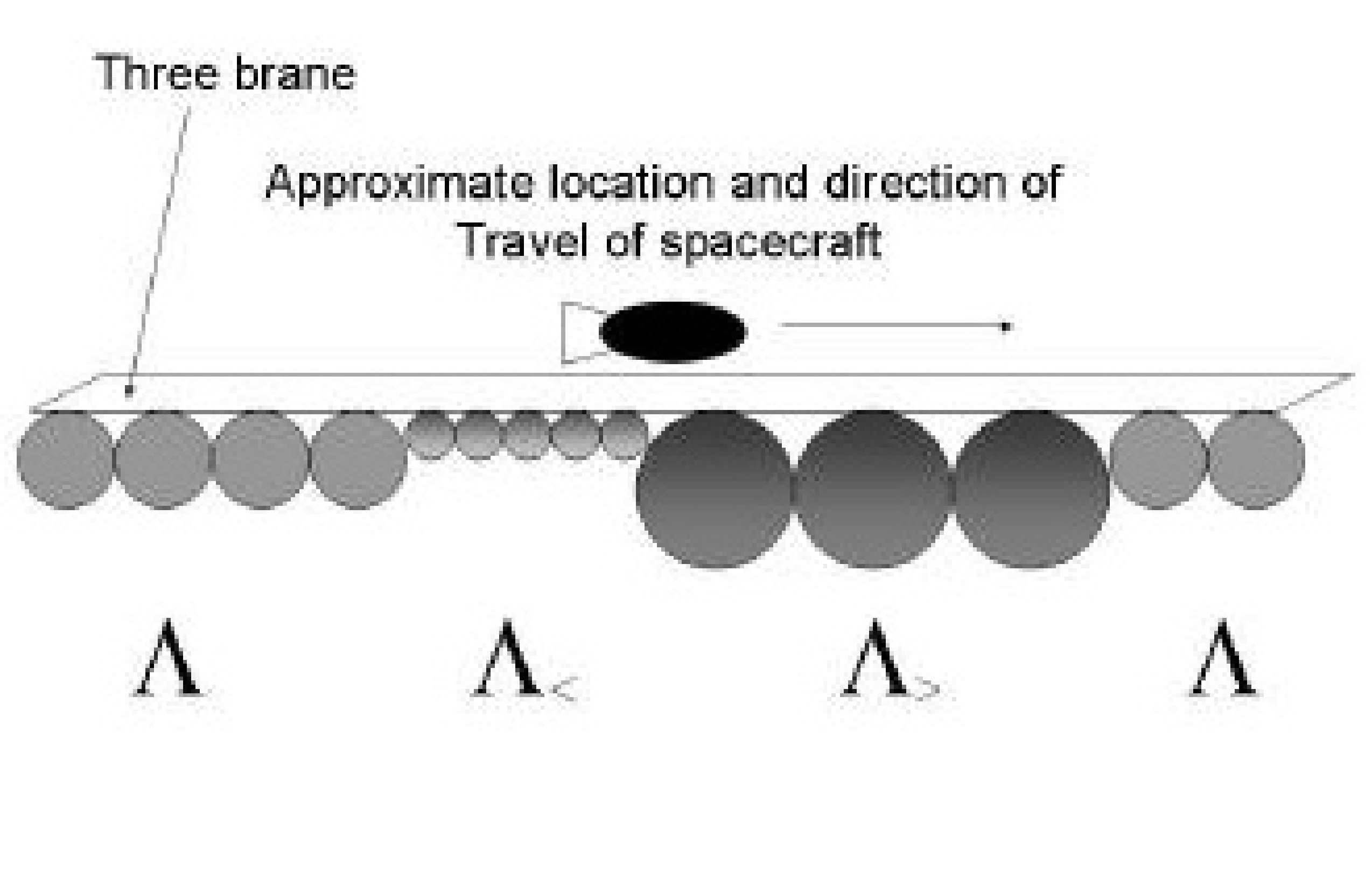}
\includegraphics[width=2.50in]{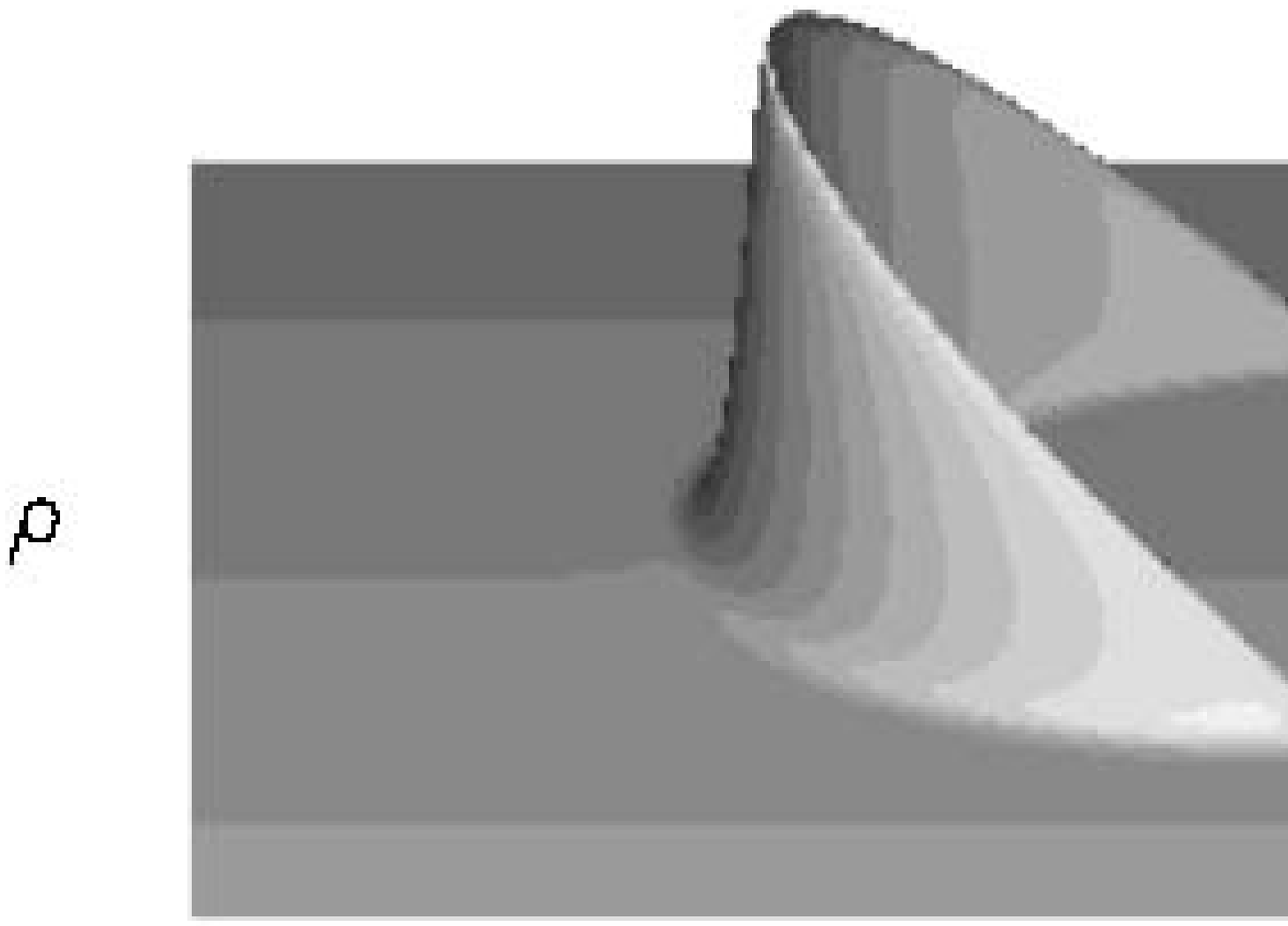}
\caption{Top image represents a spacecraft with the ability to
increase the radius of the extra dimension at it's front and
decrease the radius at its rear. The relative radius is shown
graphically as circles of smaller/larger radius when compared
against the unchanged extra dimension. The overall effect is to
locally alter the value of $\Lambda$. The Alcubierre top-hat metric
is shown beneath to demonstrate the analogous effects on spacetime.}
\end{figure}

\section{Theory}

To build the model we consider for simplicity a scalar field and its
SUSY partner a Calar field. The scalar part of the action will be
\begin{equation}S=\int d^4xd^ny\ \sqrt{\left|g
\right|}\left[\frac{1}{2}(\partial\phi)^2-\frac{1}{2}(m_0^2+\Delta
m^2\phi^2) \right]\end{equation}
where \begin{equation}\Delta
m^2(y)=m^2e^{-2{\left|y\right|}\backslash
 \delta}\end{equation} characterizes the mass-square shift and the location of the three-brane is at y=0.
The shift of the Casimir energy density due to SUSY-breaking in the
three brane 'extra' volume is [7]
\begin{eqnarray}
{\delta\rho_v(\Delta
m^2,a)}=\rho_v^{(ren)}(m^2,a)-\rho_v^{(ren)}(m^2=0 ,a) \nonumber \\
\cong \ \kappa_n\frac{a^2}{a^{4+n}}\Delta m^2(y) \ \ \ \ \ \ \ \ \ \
\ \ \ \ \ \
\end{eqnarray}

\

where $\kappa_n$ is some constant.

\

The renormalized Casimir energy density is

\begin{equation}
\rho_v^{(ren)}(m^2,a)=\rho_v(m^2,a)-\rho_v(m^2,a\to\infty)
\end{equation}

After integrating over the extra dimensional space the renormalized
Casimir energy density is

\begin{equation}
{\delta\rho_v^{(4)}} \cong
\kappa_n\frac{m^2}{a^2}\frac{\pi^{\frac{n}{2}} \ \Gamma(n)}{2^{n-1}
\ \Gamma(\frac{n}{2})}\left(\frac{\delta}{a}\right)^n
\end{equation}

where $\delta$ is the width of the brane and $a$ is the radius of
the compactified extra dimension.

The same calculations can be applied to the superpartner calar
field. The only change is a sign difference. Thus the total Casimir
energy density contribution from a scalar/calar field induced by
SUSY breaking is

\begin{eqnarray}
{\delta\rho_{total}}=\delta\rho_{scalar}(\Delta
m^2,a)+\delta\rho_{calar}(\Delta \tilde{m}^2,a)      \\
=\delta\rho_{scalar}(\Delta m^2,a)-\delta\rho_{scalar}(\Delta
\tilde{m}^2,a)
\end{eqnarray}

It is important to note that although the Casimir energy lies in the
extra dimension it does contribute to the overall energy density of
the unverse in $M^4$ and represents the $\Lambda$ term in Einstein's
equation

\begin{equation}
R_{\mu\nu}-\frac{1}{2}Rg_{\mu\nu}+\Lambda g_{\mu\nu}=8\pi G
T_{\mu\nu}\end{equation}

The important term in equation (5) is $\left({\delta}\backslash
{a}\right)^n$ which demonstrates that the Casimir energy density and
hence the cosmological constant term are immutably related to the
radius a of the extra dimension.

Consider some arbitrarily advanced civilization with the ability to
locally alter the radius of the extra dimension. This would bring
about a local shift in the cosmological constant roughly of the form

\begin{equation}
\Delta\Lambda \cong \kappa_n\frac{m^2}{a^2}\frac{\pi^{\frac{n}{2}} \
\Gamma(n)\delta^n}{2^{n-1} \ \Gamma(\frac{n}{2})} \left[
\left(\frac{1}{a}\right)^n-\left(\frac{1}{a'}\right)^n \right]
\end{equation}

where $a'$ represents the modified radius. Thus, given some
mechanism for locally adjusting the radius it is feasible that one
could locally change the value of the cosmological constant thus
providing a mechanism for pushing/pulling a spacecraft. This is
analogous to the Alcubierre bubble.
\\
\section{Bibliography}

1. M. Alcubierre 1994 Class. Quantum Grav. 11, L73

2. http://www.grc.nasa.gov/WWW/bpp/

3. http://www.greenglow.co.uk/

4. M. S. Morris, K. S. Thorne and U. Yurtsever, Wormholes, time
machines, and the weak energy conditions, Phys. Rev. Lett.,
61:1446-1449, 1988.

5. For a review see, Brane World Cosmology, Philippe Brax, Carsten
van de Bruck, Anne-Christine Davis,
Rept.Prog.Phys.67:2183-2232,2004, hep-th/0404011

6. G. F. Giudice and R. Rattazzi, Phys. Rept. 322, 419 (1999).

7. Casimir effect in a supersymmetry-breaking brane world as dark
energy. By Pisin Chen (SLAC), Je-An Gu (Taiwan, Natl. Taiwan U. and
SLAC),. SLAC-PUB-10772, Sep 2004. 4pp. e-Print Archive:
astro-ph/0409238

\end{document}